\documentclass[conference]{IEEEtran}
\IEEEoverridecommandlockouts
\usepackage{cite}
\usepackage{graphicx,color,psfrag,overpic}
\usepackage{amsmath}
\usepackage{times}
\usepackage{latexsym}
\usepackage{bm}
\usepackage{amssymb}
\usepackage{stfloats}
\usepackage{cases}
\usepackage{array}
\usepackage{setspace}
\usepackage{fancyhdr}

\usepackage{amsfonts}
\usepackage{latexsym,bm}
\usepackage{amsmath,amssymb,amsfonts}
\usepackage{indentfirst}
\usepackage{cite}
\usepackage{graphicx,color,psfrag,overpic}
\usepackage{stfloats}
\usepackage{cases}
\usepackage{array}
\usepackage{verbatim}
\usepackage{setspace}
\usepackage{fancyhdr}
\usepackage{citesort}
\usepackage{mathrsfs}

\usepackage[ruled,vlined]{algorithm2e}

\newtheorem{Proposition}{Proposition}

\usepackage{color}

    \newcommand{\qn}{{\bf n}}

    \newcommand{\qv}{{\bf v}}
    \newcommand{\qw}{{\bf w}}
    \newcommand{\qx}{{\bf x}}
    \newcommand{\qy}{{\bf y}}
    \newcommand{\qz}{{\bf z}}

    \newcommand{\qF}{{\bf F}}

    \newcommand{\qS}{{\bf S}}

    \newcommand{\qzero}{{\bf 0}}

    \newcommand{\qOmega}{{\boldsymbol \Omega}}

    \newcommand{\ty}{\tilde{y}}

    \newcommand{\tqy}{\tilde{\bf y}}

    \newcommand{\bbC}{{\mathbb C}}

    \newcommand{\calN}{{\cal N}}

    \newcommand{\calR}{{\cal R}}

    \newcommand{\Ex}{{\sf E}}
    
    \newcommand{\Varx}{{\sf Var}}
    
    \newcommand{\mmse}{{\sf mmse}}
    
    \newcommand{\sign}{{\sf sign}}

    \newcommand{\rmd}{{\rm d}}
    \newcommand{\rmD}{{\rm D}}

    \newcommand{\sfc}{{\sf c}}

    \newcommand{\sfj}{{\sf j}}

    \newcommand{\sfA}{{\sf A}}
    \newcommand{\sfB}{{\sf B}}

    \newcommand{\sfP}{{\sf P}}
    \newcommand{\sfQ}{{\sf Q}}

    \newcommand{\sfX}{{\sf X}}

    \newcommand{\scP}{\mathscr{P}}

\ifCLASSINFOpdf

\else

\fi

\begin{document}
\title{Generalized Turbo Signal Recovery for Nonlinear Measurements and Orthogonal Sensing Matrices}

\author{\IEEEauthorblockN{Ting~Liu\IEEEauthorrefmark{1}, Chao-Kai~Wen\IEEEauthorrefmark{2}, Shi~Jin\IEEEauthorrefmark{1}, and Xiaohu~You\IEEEauthorrefmark{1} }
\IEEEauthorblockA{\IEEEauthorrefmark{1}National Mobile Communications Research Laboratory,
Southeast University\\
Nanjing 210096, P. R. China,
E-mail: $\left\{\textmd{liuting, jinshi, xhyu}\right\}$@seu.edu.cn}
\IEEEauthorblockA{\IEEEauthorrefmark{2}Institute of Communications Engineering, National Sun Yat-sen University\\
Kaohsiung 804, Taiwan, E-mail: $\textmd{chaokai.wen}$@mail.nsysu.edu.tw}
}


\maketitle
\begin{abstract}
In this study, we propose a generalized turbo signal recovery algorithm to estimate a signal from  quantized measurements, in which the sensing
matrix is a row-orthogonal matrix, such as the partial discrete Fourier transform matrix.
The state evolution of the proposed algorithm is derived and is shown to
be consistent with that obtained with the replica method. Numerical experiments illustrate the excellent agreement of the proposed algorithm with theoretical
state evolution.

\end{abstract}


\maketitle
\section{Introduction}

Compressed sensing (CS) is a signal processing technique that aims to reconstruct a sparse signal with a higher-dimension $(N)$ space from an underdetermined lower-dimension $(M)$ measurement space. In the literature, $\ell_{1}$-norm minimization is the most
widely used scheme in signal reconstruction because it is convex and hence can be solved efficiently
\cite{Candes-05IT,Donoho-06IT}. Despite its efficient solution, however, $\ell_{1}$-reconstruction is far from being optimal \cite{Kabashima-09JSM}.
If the probabilistic properties of the signal are known, then the probabilistic Bayesian inference offers the optimal reconstruction in the minimum
mean-square-error (MSE) sense; however, the optimal Bayes estimation is not computationally tractable. By using belief propagation, an efficient and less complex
alternative, referred to as approximate message passing (AMP) \cite{Kabashima-03JPA,Donoho-09PNAS,Rangan-10ArXiv,Krzakala-12JSM}, has recently emerged.

The implementation of AMP still requires many matrix multiplications up to an order of $O(MN)$. Considering a special sensing matrix that allows a fast
multiplication procedure is therefore of great interest. The partial discrete Fourier transform (DFT) sensing matrix, i.e., a randomly selected DFT matrix, is one
such example \cite{Barbier-15JSM,Javanmard-12ISIT}. Using DFT as the sensing matrix, fast Fourier transform can be used to perform matrix multiplications down to
the order of $O(N\log_2N)$, and most importantly, the sensing matrix is not required to be stored. The entries of a DFT matrix are, however, not i.i.d.. In
contrast to the case with matrices with independent entries, AMP does not perform well for sensing matrices with a row-orthogonal ensemble. Recently, Ma {\em et
al.} in \cite{Ma-15SPL} developed a signal recovery algorithm for the partial DFT sensing matrix by exploiting a turbo principle in iterative decoding. In contrast
to other developments in this area, e.g., \cite{Kabashima-14ISIT,Cakmak-14ISIT}, the turbo signal recovery (turbo-SR) algorithm \cite{Ma-15SPL} has
an excellent convergence property, and most importantly, the state evolution of this algorithm perfectly agrees with that predicted by the replica method. The latter characteristic indicates the optimality of the turbo-SR algorithm over the partial DFT sensing matrix.

The turbo-SR algorithm is developed under \emph{linear} measurements. However, in CS problems, quantization is often a necessary step in the acquisition of
measurements. Especially, signal recovery problems from low-resolution (or even 1-bit) measurements are of particular interest in recent years
\cite{Boufounos-08CISS,Zymnis-10SPL,Xu-14JSM}. This study presents a novel algorithm, the \emph{generalized} turbo signal recovery (GTurbo-SR) algorithm, to
recover signal from a row-orthogonal sensing matrix followed by a quantized measurement channel. This algorithm extends the earlier turbo-SR algorithm to deal with
arbitrary distribution on the output of the measurements. The state evolution (SE) of the GTurbo-SR is derived and is shown to be consistent with that obtained
with the replica method.

{\em Notations}---For a complex-valued variable $z$, we use $z_{\rm R}$ and $z_{\rm I}$ to denote the real and imaginary parts of $z$, respectively.
A random variable $z$ drawn from the proper complex Gaussian distribution of mean $\mu$ and variance $v$ is described by
 $z \sim \calN_{\bbC}(z;\mu,v) \triangleq \frac{1}{\pi v} e^{-\frac{|z-\mu|^2}{v}}$.
We use $\rmD z$ to denote the real Gaussian integration measure
$\rmD z =  \phi(z)\, \rmd z \mbox{~~with~~} \phi(z) \triangleq \frac{1}{\sqrt{2\pi}}e^{-\frac{z^2}{2}}$,
and we use ${\rmD z_{\sfc} = \frac{e^{-|z|^2}}{\pi}} \rmd z$ to denote the complex Gaussian integration measure. Finally,
    $\Phi(x) \triangleq \frac{1}{\sqrt{2\pi}} \int_{-\infty}^x e^{-\frac{t^2}{2}} \, \rmd t $
denotes the cumulative Gaussian distribution function, and we have ${\partial \Phi(x)/\partial y = \phi(x) \partial x /\partial y}$.

\section{Problem Formulation}

\begin{figure*}
\begin{center}
\resizebox{4.75in}{!}{%
\includegraphics*{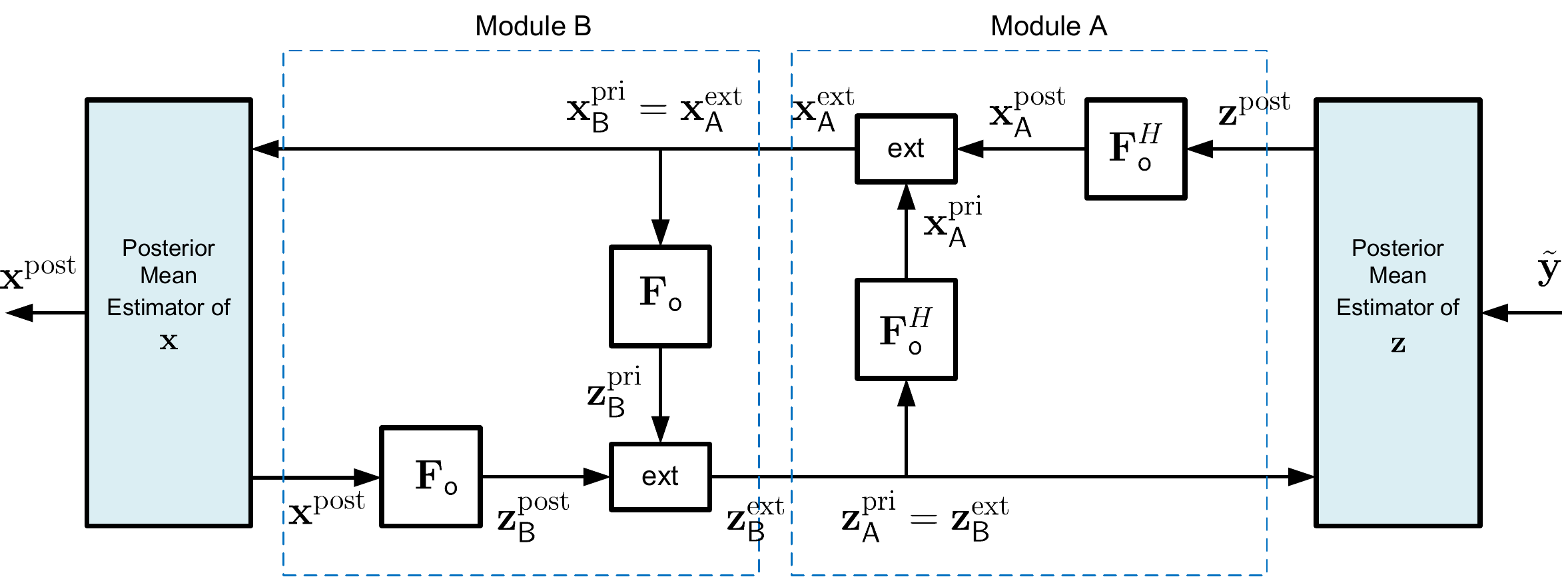} }%
\caption{Block diagram of the GTurbo-SR algorithm.}\label{fig:GTurbo-SR}
\end{center}
\end{figure*}

We consider the noisy CS problem
\begin{equation}\label{eq:sysModel}
    \qy= \qF \qx + \qn,
\end{equation}
where $\qy \in \bbC^{M}$ is a measurement vector, $\qF \in \bbC^{M \times N}$ denotes a known sensing matrix, $\qx \in \bbC^{N}$ is a signal vector, and $\qn
\in\bbC^{M}$ is the additive white Gaussian noise vector with zero mean and element-wise variance $\sigma^2$. We denote by $\alpha = M/N$ the measurement ratio
(i.e., number of measurements per variable). The sensing matrix $\qF$ is obtained by random selection of a set of rows from an unitary matrix or the standard
DFT matrix $\qF_{\sf o} = [\frac{ e^{-2\pi \sfj (m-1)(n-1)/N}}{\sqrt{N}} ] \in \bbC^{N \times N}$. We refer to such $\qF$ as the row-orthogonal matrix.

For ease of ``expression'' and convenience, we work with the enlarged orthogonal matrix with rows and columns by setting them to zero rather than removing them. Therefore, we
enlarge $\qy$ and $\qw$ to be $N$-dimensional vectors by zero padding and denote them by $\underline{\qy} \in \bbC^{N}$ and $\underline{\qn} \in \bbC^{N}$. We let
matrix $\qS$ be a diagonal projection matrix, in which its off-diagonal entries are all zeros, and its diagonal entries are zeros or ones. Let
\begin{equation} \label{eq:z=Fx}
    \qz = \qF_{\sf o}\qx,
\end{equation}
be an $N$-dimensional vector. Then, the input output relationship of (\ref{eq:sysModel}) can be equivalently expressed as
\begin{equation}\label{eq:sysModel_Enlarge}
    \underline{\qy} = \qS \qz + \underline{\qn}.
\end{equation}
Note that all the following descriptions are based on the enlarged system (\ref{eq:sysModel_Enlarge}). By abuse of notation, we continue to write $\qy$ and $\qn$ for $\underline{\qy}$ and $\underline{\qn}$, respectively.

In this study, we are interested in the measurements acquired through quantizers. Let $\sfQ_{c}(\cdot)$ be a complex-valued quantizer $\sfQ_{c}(\cdot)$,
which is defined as $\ty_{n} = \sfQ_{c}(y_{n}) \triangleq \sfQ(y_{{\rm R},n}) + \sfj\sfQ(y_{{\rm I},n} )$, i.e., the real and imaginary parts are quantized
separately.
The quantization is applied element-wise on each measurement, and the resulting quantized signal $\tqy$ is given by
\begin{equation} \label{eq:qsys}
    \tqy = \sfQ_{c}{\left(\qy\right)}.
\end{equation}
The output is assigned the value $\ty_{n}$ when the quantizer input falls in the interval $(\ty_{n}^{{\rm low}}, \,\ty_{n}^{{\rm up}}]$. \footnote{Here, $\ty_{n}$ should be specified as $\ty_{{\rm R},n}$ or $\ty_{{\rm I},n}$. We abuse $\ty_{n}$ to denote each real channel.} For example, for a typical uniform $\sfB$-bit
quantizer with quantization step size $\Delta$, the quantized output is given by
\begin{equation}
    \ty_{n} \in \left\{ {\left( -\frac{1}{2} + b \right)}\Delta;~ b= -\frac{2^\sfB}{2}+1, \cdots,  \frac{2^\sfB}{2}  \right\} \triangleq \calR_{\sfB},
\end{equation}
and the associated lower and upper thresholds are given by
\begin{subequations}\label{quanbound}
\begin{align}
    \ty_{n}^{\rm low} &= \left\{
    \begin{array}{ll}
    \ty_{n}-\frac{\Delta}{2}, & \textrm{for $\ty_{n}\ge -{\left(\frac{2^{\sfB}}{2}-1\right)}\Delta$},\\
    -\infty, & \textrm{otherwise},
    \end{array} \right. \\
    \ty_{n}^{\rm up} &= \left\{ \begin{array}{ll}
    \ty_{n}+\frac{\Delta}{2}, & \textrm{for $\ty_{n}\le {\left(\frac{2^{\sfB}}{2}-1\right)}\Delta$},\\
    -\infty, & \textrm{otherwise}.
    \end{array} \right.
\end{align}
\end{subequations}

Our aim is to estimate $\qx$ and $\qz$ from $\tqy$ given $\qF$ (or, more precisely, $\qS\qF_{\sf o}$). We consider the Bayesian optimal inference because this
methodology can achieve the best estimates in terms of MSE. Toward this end, we introduce the distributions of signals $\qx$ and measurements
$\tqy$. We suppose that each entry of $\qx$ is generated from a distribution $\sfP_{\sfX}(x)$ independently,
\begin{equation}\label{eq:px}
    \sfP_{\sfX}(\qx) = \prod_{n=1}^{N} \sfP_{\sfX}(x_n) .
\end{equation}
On the other hand, the distribution of the quantized measurements under (\ref{eq:qsys}), conditional on $\qz$, is given by
\begin{equation} \label{eq:lnkelihood}
    \sfP_{\sf out}(\tqy|\qz) = \prod_{n=1}^{N} {\sfP_{\sf out}{( \ty_{n} | z_{n})}},
\end{equation}
where
\begin{equation} \label{eq:lnkelihood_each}
    \sfP_{\sf out}{( \ty_{n} | z_{n} )}
    = \left\{
    \begin{array}{ll}
    1, & \mbox{if }  S_{nn} = 0, \\
    \Psi{\left(\ty_{{\rm R},n};z_{{\rm R},n}, \frac{\sigma^2}{2} \right)} \\
    \times \Psi{\left(\ty_{{\rm I},n};z_{{\rm I},n}, \frac{\sigma^2}{2} \right)}, & \mbox{if }  S_{nn} = 1,
    \end{array}
    \right.
\end{equation}
with
\begin{equation} \label{eq:def_Psi_b}
    \Psi{\left(\ty;z, c^2\right)} \triangleq \Phi{\left( \frac{\ty^{\rm up}- z}{c}\right)} - \Phi{\left( \frac{\ty^{\rm low}- z}{c}\right)}.
\end{equation}



\section{Generalized Turbo Signal Recovery }

The GTurbo-SR is presented in Algorithm \ref{ago:GTurbo-SR}, which is a generalization of the turbo-SR algorithm in \cite{Ma-15SPL}.
Figure \ref{fig:GTurbo-SR} illustrates the block diagram of the algorithm. The idea of the algorithm use the turbo principle in iterative decoding to compute extrinsic messages of $\qx$ and $\qz$ in
modules A and B, on the basis of \emph{linear} transform $\qz = \qF_{\sf o} \qx$. Specifically, in lines 3 and 4, $\qx_{\sf A}^{\rm post}$ and $v_{\sf A}^{\rm post}$ are estimates of
$\qx = \qF_{\sf o}^{H} \qz$ and its corresponding variance, respectively. Subsequently, by excluding the prior knowledge $(\qx_{\sf A}^{\rm pri},
v_{\sf A}^{\rm pri})$ from the posterior moments $(\qx_{\sf A}^{\rm post}, v_{\sf A}^{\rm post})$, lines 5--7 compute the extrinsic mean and variance of $\qx$.
Similar to those in lines 3 and 4, $\qz_{\sf B}^{\rm post}$ and $v_{\sf B}^{\rm post}$ in lines 10 and 11 are estimates of $\qz = \qF_{\sf o} \qx$ and its corresponding variance, respectively.
The extrinsic mean and variance of $\qz$ are then evaluated in lines 12--14.

In the GTurbo-SR algorithm, the estimate of the posterior mean and the variance of $\qz$ in lines 1 and 2 are nonlinear, and they consider the prior mean and
variance of $\qz = \qF_{\sf o}\qx$ from $\qz_{\sfA}^{\rm pri}$ in line 12 and $v_{\sfA}^{\rm pri}$ in line 13. Assuming the prior ${\sfP_{\sf Z}( z_{n} )} =
\calN_{\bbC}(z; z_{{\sf A},n}^{\rm pri}, v_{\sfA}^{\rm pri})$ and combining $ \sfP_{\sf out} ( \ty_{n} | z_{n} )$ in (\ref{eq:lnkelihood_each}), we take the expectation
and variance in lines 1 and 2 with respect to (w.r.t.) the posterior probability
\begin{equation*} 
    \scP(z_{n}) = \frac{{\sfP_{\sf out}( \ty_{n} | z_{n} )} {\sfP_{\sf Z}( z_{n} )}}
    { \int {\sfP_{\sf out}( \ty_{n} | z'_{n} )} \sfP_{\sf Z}( z'_{n} )\,\rmd z'_{n} }.
\end{equation*}
Similarly, the posterior mean and variance of $\qx$ in lines 8 and 9 are taken w.r.t. the posterior probability
\begin{align}
    \scP(x_{n}) &=
    \frac{\calN_{\bbC}(x_{n}; \qx_{\sf B}^{\rm pri}, v_{\sf B}^{\rm pri}) {\sfP_{\sfX}( x_{n} )}}
    { \int \calN_{\bbC}(x'_{n}; \qx_{\sf B}^{\rm pri}, v_{\sf B}^{\rm pri}) {\sfP_{\sfX}( x'_{n} )}\,\rmd x'_{n}}. \label{eq:margPost_x2}
\end{align}
The algorithm produces a sequence of estimates $\qx^{\rm post}$ and $\qz^{\rm post}$ for the unknown vectors $\qx$ and $\qz$, respectively.

\begin{algorithm}[!h]\label{ago:GTurbo-SR}  \footnotesize
\scriptsize
  \caption{GTurbo-SR}
  \SetKwInOut{Input}{input}
  \SetKwInOut{Output}{output}
  \SetKwInOut{Initialize}{initialize}
  \SetKwInOut{Definition}{definition}


  \Input{Quantized observations $\tqy$, sensing matrix $\qF$, likelihood $\sfP_{\sf out}( \ty | z )$, and prior distributions $\sfP_{\sfX}(\qx)$}
  \BlankLine
  \Output{Recovered signal $\hat{\qx}$ }
  \BlankLine
  \Initialize{$t \leftarrow 1$, $\qz_{\sfA}^{\rm pri} \leftarrow \qzero$, and $v_{\sfA}^{\rm pri} \leftarrow \Ex\{ |x|^2 \}$
  }
  \BlankLine
  \vspace{0.3cm}
  \While{$t < T_{\max}$ }{
  1) {\bf Output nonlinear steps}: \\
  \hspace{0.25cm} Compute the posterior mean and variance of $\qz$ \\
  \nl \hspace{0.5cm} $\qz^{\rm post} \leftarrow  \Ex{\left\{ \qz \big| \qz_{\sf A}^{\rm pri}, v_{\sfA}^{\rm pri} \right\}}$\;
  \nl \hspace{0.5cm} $\qv_{z}^{\rm post} \leftarrow  \Varx{\left\{ \qz \big| \qz_{\sf A}^{\rm pri}, v_{\sfA}^{\rm pri} \right\}}$\;
  \BlankLine
  2) Estimate $\qx$ via $\qF_{\sf o}^{H}\qz$ and its corresponding variance \\
  \nl \hspace{0.5cm} $\qx_{\sf A}^{\rm post} \leftarrow  \qF_{\sf o}^{H} \qz^{\rm post} $\;
  \nl \hspace{0.5cm} $v_{\sf A}^{\rm post} \leftarrow  \frac{1}{N}\sum_{n=1}^{N} v_{z, n}^{\rm post}$\;
  \BlankLine
  \hspace{0.25cm} Compute the extrinsic mean and variance of $\qx$ \\
  \nl \hspace{0.5cm} $\qx_{\sf A}^{\rm pri} \leftarrow \qF_{\sf o}^{H} \qz_{\sfA}^{\rm pri} $ \;
  \nl \hspace{0.5cm} $v_{\sf B}^{\rm pri} \leftarrow v_{\sf A}^{\rm ext} \leftarrow  \left( \frac{1}{v_{\sf A}^{\rm post}} - \frac{1}{v_{\sf A}^{\rm pri}} \right)^{-1} $\;
  \nl \hspace{0.5cm} $\qx_{\sf B}^{\rm pri} \leftarrow \qx_{\sf A}^{\rm ext} \leftarrow  v_{\sf A}^{\rm ext} \left( \frac{\qx_{\sf A}^{\rm post}}{v_{\sf A}^{\rm post}} - \frac{\qx_{\sf A}^{\rm pri}}{v_{\sf A}^{\rm pri}} \right)$\;
  \BlankLine
  3) {\bf Input nonlinear steps}: \\
  \hspace{0.25cm} Compute the posterior mean and variance of $\qx$ \\
  \nl \hspace{0.5cm} $\qx^{\rm post} \leftarrow  \Ex{\left\{ \qx \big| \qx_{\sf B}^{\rm pri}, v_{\sf B}^{\rm pri} \right\}}$\;
  \nl \hspace{0.5cm} $\qv_{x}^{\rm post} \leftarrow  \Varx{\left\{ \qx \big| \qx_{\sf B}^{\rm pri}, v_{\sf B}^{\rm pri} \right\}}$\;
  \BlankLine
  4) Estimate $\qz$ via $\qF_{\sf o}\qx$ and its corresponding variance \\
  \nl \hspace{0.5cm} $\qz_{\sf B}^{\rm post} \leftarrow  \qF_{\sf o} \qx^{\rm post} $\;
  \nl \hspace{0.5cm} $v_{\sf B}^{\rm post} \leftarrow  \frac{1}{N}\sum_{n=1}^{N} v_{x,n}^{\rm post}$\;
  \BlankLine
  \hspace{0.25cm} Compute the extrinsic mean and variance of $\qz$ \\
  \nl \hspace{0.5cm}  $\qz_{\sf B}^{\rm pri} \leftarrow \qF_{\sf o} \qx_{\sf B}^{\rm pri} $ \;
  \nl \hspace{0.5cm} $v_{\sf A}^{\rm pri} \leftarrow v_{\sf B}^{\rm ext} \leftarrow  \left( \frac{1}{v_{\sf B}^{\rm post}} - \frac{1}{v_{\sf B}^{\rm pri}} \right)^{-1} $\;
  \nl \hspace{0.5cm} $\qz_{\sf A}^{\rm pri} \leftarrow \qz_{\sf B}^{\rm ext} \leftarrow  v_{\sf B}^{\rm ext} \left( \frac{\qz_{\sf B}^{\rm post}}{v_{\sf B}^{\rm post}} - \frac{\qz_{\sf B}^{\rm pri}}{v_{\sf B}^{\rm pri}} \right)$\;
  \BlankLine
  $t \leftarrow t+1$ \;
  }
\end{algorithm}

Explicit expressions of the posterior mean and variance of $z_{n}$ are provided in \cite{Wen-15TSP} and are shown in (\ref{eq:hatZ_RealGaussian}) and
(\ref{eq:mseZ_RealGaussian}) at the top of the next page.
\begin{figure*}
\normalsize
\begin{align}
    z_{n}^{\rm post}
   &= \left\{
   \begin{aligned}
   &z_{{\sf A},n}^{\rm pri}, & \mbox{if } S_{nn} = 0, \\
   &z_{{\sf A},n}^{\rm pri} + \frac{\sign(\ty_{n}) v_{\sfA}^{\rm pri} }{\sqrt{2(\sigma^2 + v_{\sfA}^{\rm pri})}} \left( \frac{\phi(\eta_1)-\phi(\eta_2)}{\Phi(\eta_1)-\Phi(\eta_2)} \right),
   & \mbox{if } S_{nn} = 1,
   \end{aligned}
   \right.
   \label{eq:hatZ_RealGaussian} \\
    v_{z,n}^{\rm post}
      &= \left\{
   \begin{aligned}
   &\frac{v_{\sfA}^{\rm pri}}{2}, & \mbox{if } S_{nn} = 0, \\
   &\frac{v_{\sfA}^{\rm pri}}{2} - \frac{(v_{\sfA}^{\rm pri})^2}{2(\sigma^2 + v_{\sfA}^{\rm pri})}\times
     \left( \frac{\eta_1\phi(\eta_1)-\eta_2\phi(\eta_2)}{\Phi(\eta_1)-\Phi(\eta_2)}
     + \left(\frac{\phi(\eta_1)-\phi(\eta_2)}{\Phi(\eta_1)-\Phi(\eta_2)}\right)^2 \right),
     & \mbox{if } S_{nn} = 1,
      \end{aligned}
   \right.
     \label{eq:mseZ_RealGaussian}
\end{align}
where
\begin{equation} \label{eq:eta_def}
    \eta_1 = \frac{\sign(\ty_{n})z_{{\sf A},n}^{\rm pri}-\min\{|\ty_{n}^{\rm low}|,|\ty_{n}^{\rm up}|\}}{\sqrt{(\sigma^2 + v_{\sfA}^{\rm pri})/2}}, ~~~
    \eta_2 = \frac{\sign(\ty_{n})z_{{\sf A},n}^{\rm pri}-\max\{|\ty_{n}^{\rm low}|,|\ty_{n}^{\rm up}|\}}{\sqrt{(\sigma^2 + v_{\sfA}^{\rm pri})/2}}.
\end{equation}
\hrulefill
\end{figure*}
We have
abused $\ty_{n}$ and $z_{n}^{\rm post}$ in (\ref{eq:hatZ_RealGaussian}) and (\ref{eq:mseZ_RealGaussian}) to denote $\ty_{{\rm R},n}$ and $z_{{\rm R},n}^{\rm
post}$, respectively. The estimator for the imaginary part $z_{{\rm I},n}^{\rm post}$ can be obtained analogously as (\ref{eq:hatZ_RealGaussian}) and
(\ref{eq:mseZ_RealGaussian}), whereas $\ty_{n}$ should be replaced by $\ty_{{\rm I},n}$.

On the other hand, suppose that the elements of $\qx$ are drawn i.i.d. from the Bernoulli-Gaussian (BG) distribution, i.e.,
\begin{equation}
\sfP_{\sf X}(x) = (1-\rho)\delta(x) + \rho \calN_{\bbC}(x;0,\varsigma_{x}),
\end{equation}
where $\delta(\cdot)$ denotes the Dirac delta, $\rho$ the sparsity rate, and $\varsigma_{x}$ the active-coefficient variance.
Then, the explicit expressions of the posterior mean and variance of $\qx$ in lines 8 and 9 are given by \cite{Barbier-15JSM}.
Please refer also to \cite{Wen-15TSP} for other distributions of $\sfP_{\sfX}(x_n)$ such as the QAM constellations.

\section{State Evolution}

The behavior of the GTurbo-SR algorithm can be described by a set of SE equations. We then derive the SE equations in the large-system regime where $M$ and $N$ reach
infinity, whereas the ratio $M/N = \alpha$ remains fixed.

Similar to the notation used in \cite{Ma-15SPL}, we define
\begin{equation}
 \eta = 1/v_{\sf B}^{\rm pri}~~ \mbox{and}~~ v = v_{\sf A}^{\rm pri}.
\end{equation}
Let us consider a scalar AWGN channel
\begin{equation} \label{eq:ScalarSysModel}
    r = x +  w,
\end{equation}
where $w \sim \calN_{\bbC}(w; 0, \eta^{-1})$.
The posterior mean estimator of $x$ from (\ref{eq:ScalarSysModel}) is given by
\begin{equation} \label{eq:ScalarEstX}
    \Ex\{x|r\} = \int  x \sfP(x|r) \rmd x,
\end{equation}
where $\sfP(x|r) = \sfP(r|x) \sfP_{\sfX}(x)/\sfP(r)$ and $\sfP(r|x) =\frac{\eta}{\pi} e^{-\eta |r-x|^2}$. Then, we define $\mmse(\cdot)$ of this estimator as
\begin{equation} \label{eq:defMMSE}
    \mmse(\eta) \triangleq \Ex\left\{ \left|x-\Ex\{x|r\}\right|^2 \right\},
\end{equation}
where the expectation is taken over the joint distribution $\sfP(r,x) = \sfP(r|x) \sfP_{\sfX}(x)$.

Combining the above definitions into lines 9, 11, and 13 of Algorithm \ref{ago:GTurbo-SR}, we can easily characterize the SE of $v_{\sf A}^{\rm pri}$ as
\begin{equation}
     v^{t+1} = \left( \frac{1}{\mmse(\eta^{t+1})} - \eta^{t+1} \right)^{-1},
\end{equation}
where the superscript $t$ represents the iteration indices. Next, with $v^{t+1}$, we aim at evaluating $v_{\sf A}^{\rm post}$ in line 4 of Algorithm
\ref{ago:GTurbo-SR}. Toward this end, we have to obtain the large-system behavior of $v_{\sf A}^{\rm post}$. In the large-system limit
$v_{\sf A}^{\rm post}$ for a given $\tqy$ has small deviations from the expectation of $v_{z,n}^{\rm post}$ w.r.t. $\tqy$, which is called the
self-averaging property in the large-system limit.
To compute this expectation, we need the joint distribution $\sfP(\tqy,\qz_{\sf A}^{\rm pri})$. Our strategy to obtain $\sfP(\tqy,\qz_{\sf A}^{\rm pri})$ is via the
marginal distribution $\sfP(\tqy,\qz_{\sf A}^{\rm pri}, \qz) = \sfP(\tqy|\qz_{\sf A}^{\rm pri}, \qz) \sfP(\qz_{\sf A}^{\rm pri}, \qz)$.

First we consider the joint
distribution $\sfP(\qz_{\sf A}^{\rm pri}, \qz)$. We notice that both $\qz_{\sf A}^{\rm pri}$ and $\qz$ are sums over many independent terms. Therefore, according
to the CLT, they can be approximated as Gaussian random variables. Their means are zero because $\{ x_{n} \}$ are zero means. The covariance
matrix between $z_{n}$ and $z_{{\sf A},n}^{\rm pri}$ can be shown to be
\begin{equation}
    \qOmega = \left[
    \begin{array}{cc}
    v_{x} & v_{x} - v_{\sfA}^{\rm pri} \\
    v_{x} - v_{\sfA}^{\rm pri} & v_{x} - v_{\sfA}^{\rm pri}
    \end{array}
    \right],
\end{equation}
where $v_{x} \triangleq \Ex\{ |x|^2 \}$ is the variance of $x$.
We find that the covariance matrix becomes asymptotically independent of
index $n$. Therefore, we omit index $n$ from the
covariance matrix between $z_{n}$ and $z_{{\sf A},n}^{\rm pri}$.
Altogether, these provide the following bivariate Gaussian distribution:
\begin{equation}\label{eq:biGau}
\sfP(z_{\sf A}^{\rm pri}, z)
=\underbrace{\calN\Big(z; z_{\sf A}^{\rm pri}, v_{\sfA}^{\rm pri} \Big)}_{=\sfP(z|z_{\sf A}^{\rm pri})}
 \underbrace{\calN\Big(z_{\sf A}^{\rm pri};0, v_{x} - v_{\sfA}^{\rm pri} \Big)}_{=\sfP(z_{\sf A}^{\rm pri})}.
\end{equation}

Next,
we have $\sfP(\tqy|\qz_{\sf A}^{\rm pri}, \qz) = \sfP(\tqy|\qz)$, and, thus, we obtain
\begin{equation} \label{eq:ycz}
\sfP(\ty_{n}|z_n) = \int_{\ty^{\rm low}}^{\ty^{\rm up}} \calN\left(y; z, \frac{\sigma^2}{2} \right) {\rm d} y.
\end{equation}
Note that for ease of explanation, we consider the real parts of $\ty_{n}$ and $z_n$ in (\ref{eq:ycz}). Therefore, the power of the corresponding parameters in the subsequent
expressions should be half per real and imaginary part. Combining (\ref{eq:biGau}) and (\ref{eq:ycz}) and using the definition in (\ref{eq:def_Psi_b}), we obtain
\begin{align}
&\sfP(\ty_{n}, z_{{\sf A},n}^{\rm pri}) \nonumber \\
=& \int_{-\infty}^{\infty} \sfP(\ty_{n}, z_{{\sf A},n}^{\rm pri}, z_{n}) {\rm d} z_{n} \nonumber \\
=& \int_{-\infty}^{\infty}  \int_{\ty_{n}^{\rm low}}^{\ty_{n}^{\rm up}}  \calN{\Big(y; z, \frac{\sigma^2}{2} \Big)}
     \nonumber \\
  & \hspace{1cm}  \times \calN{\Big(z; z_{\sf A}^{\rm pri}, \frac{v_{\sfA}^{\rm pri}}{2} \Big)}
  \calN{\Big(z_{\sf A}^{\rm pri}; 0, \frac{v_{x}}{2} - \frac{v_{\sfA}^{\rm pri}}{2} \Big)}
  {\rm d} y\,{\rm d} z
      \nonumber \\
=&
\Psi{\Big(\ty_{n};z_{\sf A}^{\rm pri},\frac{\sigma^2}{2}+\frac{v_{\sfA}^{\rm pri}}{2} \Big)}
\calN{\Big(z_{\sf A}^{\rm pri}; 0, \frac{v_{x}}{2} - \frac{v_{\sfA}^{\rm pri}}{2} \Big)}. \label{eq:p_yz}
\end{align}

Using (\ref{eq:p_yz}), we can evaluate the expectation of $v_{{z},n}^{\rm post}$ in (\ref{eq:mseZ_RealGaussian}) w.r.t. $(\ty_{n},z_{{\sf A},n}^{\rm
pri})$. To perform this expectation, we write $v_{{z},n}^{\rm post}$ in (\ref{eq:mseZ_RealGaussian}) for $S_{nn}=1$ as
\begin{multline}
   v_{{z},n}^{\rm post} = \frac{v_{\sfA}^{\rm pri}}{2} -  \left(\frac{v_{\sfA}^{\rm pri}}{2}\right)^2
   \left(
   \frac{ \Psi''{\Big(\ty_n;z_{{\sf A},n}^{\rm pri},\frac{\sigma^2}{2}+\frac{v_{\sfA}^{\rm pri}}{2} \Big)} }
  {\Psi{\Big(\ty_n;z_{{\sf A},n}^{\rm pri},\frac{\sigma^2}{2}+\frac{v_{\sfA}^{\rm pri}}{2} \Big)}}
  \right.
   \\
   \left.
  +
   \left( \frac{ \Psi'{\Big(\ty_n;z_{{\sf A},n}^{\rm pri},\frac{\sigma^2}{2}+\frac{v_{\sfA}^{\rm pri}}{2} \Big)} }
  {\Psi{\Big(\ty_n;z_{{\sf A},n}^{\rm pri},\frac{\sigma^2}{2}+\frac{v_{\sfA}^{\rm pri}}{2} \Big)}} \right)^2 \right),
\end{multline}
where the definition of $\Psi(\cdot)$ is given by (\ref{eq:def_Psi_b}), and
\begin{equation}  \label{eq:Psip_def}
\Psi'(\ty;z,c^2)\triangleq \frac{\partial \Psi(\ty;z,c^2)}{\partial z}
 =  - \frac{ e^{-\frac{(\ty^{\rm up}-z)^2}{2c^2}}
-e^{-\frac{(\ty^{\rm low}-z)^2}{2c^2}}}{\sqrt{2 \pi c^2}}.
\end{equation}
As a result, the expectation of $v_{{z},n}^{\rm post}$ reads
\begin{multline} \label{eq:vA_post}
 v_{\sfA}^{\rm post} = v_{\sfA}^{\rm post}
  - \alpha (v_{\sfA}^{\rm post})^2 \\
  \times \left(\sum_{\ty \in \calR_{\sfB}} \int\!\rmD z \frac{\Big[\Psi'{\Big(\ty;\sqrt{\frac{v_{x}}{2} - \frac{v_{\sfA}^{\rm pri}}{2}}z,\frac{\sigma^2}{2}+\frac{v_{\sfA}^{\rm pri}}{2} \Big)}\Big]^2}
  {\Psi{\Big(\ty;\sqrt{\frac{v_{x}}{2} - \frac{v_{\sfA}^{\rm pri}}{2}}z,\frac{\sigma^2}{2}+\frac{v_{\sfA}^{\rm pri}}{2} \Big)}} \right).
\end{multline}
Note that we have included the
contributions of the real and imaginary parts of $v_{{z},n}^{\rm post}$ into (\ref{eq:vA_post}). Substituting (\ref{eq:vA_post}) into line 6 of Algorithm
\ref{ago:GTurbo-SR}, we have closed the entire loop and present the following proposition.

\begin{Proposition}
The SE of the GTurbo-CS algorithm is characterized by
\begin{align}
     \vartheta^{t} &= \sum_{\ty \in \calR_{\sfB}} \int\! \rmD z
     \frac{ \left[ \Psi'{\Big(\ty;\sqrt{\frac{v_{x}}{2} - \frac{v^{t}}{2}} z,\frac{\sigma^2}{2}+\frac{v^{t}}{2} \Big)} \right]^2 }
     { \Psi{\Big(\ty;\sqrt{\frac{v_{x}}{2} - \frac{v^{t}}{2}} z,\frac{\sigma^2}{2}+\frac{v^{t}}{2} \Big)} },  \label{eq:SEa} \\
     \eta^{t+1} &= \frac{1}{ (\alpha \vartheta^{t})^{-1} - v^{t}}, \label{eq:SEb}  \\
        v^{t+1} &= \left( \frac{1}{\mmse(\eta^{t+1})} - \eta^{t+1} \right)^{-1}, \label{eq:SEc}
\end{align}
with initialization $v^{0} = v_{x} = \Ex\{ |x|^2 \}$.
\end{Proposition}

Interestingly, as $t \rightarrow \infty$, the SE converges to the SE equations derived from the replica method
\cite{Shinzato-09JPA}.


\section{Numerical Examples}

\begin{figure}
\begin{center}
\resizebox{2.9in}{!}{%
\includegraphics*{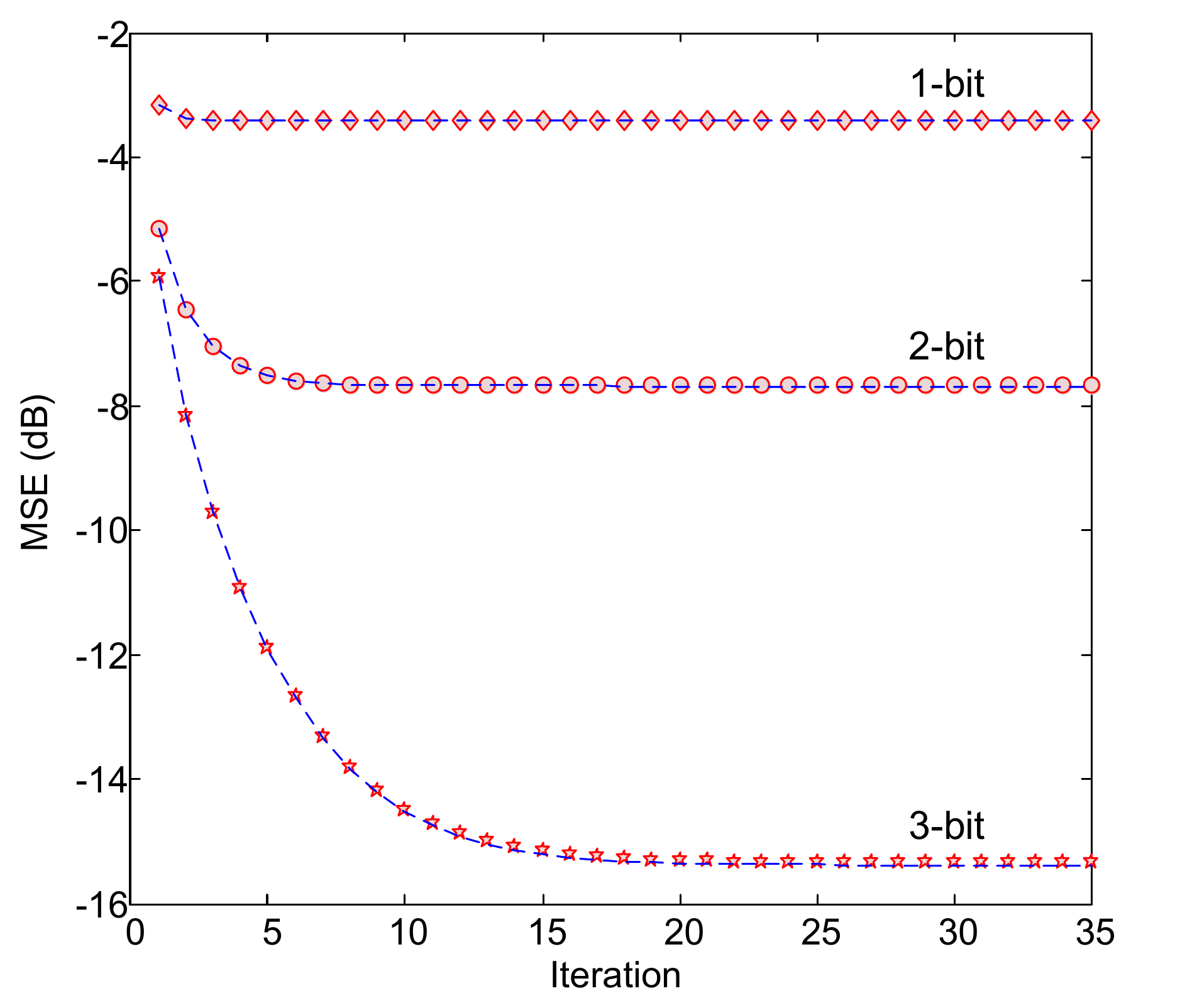} }%
\caption{MSE results under different quantization levels. $N=8192$, $M=5734 (\approx 0.7N)$, and ${\rm SNR} = 50$\,dB.
The markers denote the simulation results, whereas
the curves characterize the analytical behavior.}\label{fig:MSE_Gau_N8192}
\end{center}
\end{figure}

Computer simulations are conducted to verify the accuracy of our analytical results. In particular, we compare the SE of $\mmse(\eta^{t+1})$ in (\ref{eq:SEc}) with those obtained by the simulations.
In each simulation,
the sensing matrix is generated from a randomly scrambled $N \times N$ DFT matrix, and we perform Algorithm \ref{ago:GTurbo-SR} and compute the MSE
\begin{equation}
    {\rm MSE} = \|\qx-\qx^{\rm post}\|^2/N.
\end{equation}
The simulation results are obtained by averaging of over 2,000 realizations.
The SNR is defined as ${\rm SNR} = 1/\sigma^2$.

For comparison, the simulation scenarios completely follow those presented in \cite{Ma-15SPL}. The system parameters
are set as follows: $\alpha = 0.7$, $N=8192$, $M=5734 (\approx 0.7N)$, and ${\rm SNR} = 50$\,dB. The signal distribution $\sfP_{\sfX}(x)$ follows BG distribution with $\rho=0.4$ and $\varsigma_{x}=1/\rho$.
In this case, $\mmse$ in (\ref{eq:defMMSE}) can be obtained explicitly \cite{Tulino-13IT}
\begin{multline} \label{eq:defMMSE2}
    \mmse(\eta) = \rho \varsigma_{x} - \frac{{{{(\rho {\varsigma _x})}^2}\eta }}{{\eta {\varsigma _x} + 1}} \\
    \times \int \rmD z_{\sfc} \frac{{|z{|^2}}}{{\rho  + (1 - \rho ){e^{ - |z{|^2}\eta {\varsigma _x}}}(\eta {\varsigma _x} + 1)}}.
\end{multline}
We use the typical uniform quantizer with quantization step size $\Delta = 2^{1-\sfB}$.
Figure \ref{fig:MSE_Gau_N8192} shows the corresponding MSE results under different quantization levels. The markers denote the simulation results, whereas the curves characterize the analytical behavior.
The figure clearly demonstrates that the SE analysis precisely predicts the per iteration performance.

Other simulations, which are not reported here, show that the GTurbo-SR algorithm as well as the SE analysis, can also apply to other priors and other nonlinear measurements. For arbitrary distributions on the measurements $\sfP_{\sf out}(\ty|z)$, $\Psi(\cdot)$ in
our study should be replaced with
\begin{equation}
    \int \rmD u \, \sfP_{\sf out}{\left(\ty \left| \sqrt{\frac{v^{t}}{2}} u + \sqrt{\frac{v_{x}}{2} - \frac{v^{t}}{2}} z \right. \right)}.
\end{equation}

\section{Conclusion}

We presented a novel algorithm called GTurbo-SR algorithm to estimate a signal vector observed through a row-orthogonal sensing matrix followed by quantized
measurements. The SE analysis was provided to precisely describe the asymptotic behavior of the GTurbo-SR algorithm. The GTurbo-SR algorithm can be
applied to a large class of nonlinear measurements with a precisely predicable asymptotic SE. Other related works with random orthogonal ensembles can be found in \cite{Opper-16Arxiv,J.Ma-16Arxiv}.


%


\end{document}